\documentclass[aps,prb,floats,reprint,superscriptaddress]{revtex4-1}
\usepackage{amssymb}
\usepackage{amsmath}
\usepackage{graphicx}

 \newcommand{\const}{\mbox{const}}
 \def\bc{\begin{center}}          \def\ec{\end{center}}
 \allowdisplaybreaks
\begin{document}
 \title{Modification of narrow ablating capillaries under the influence of multiple femtosecond laser pulses}
 \author{K.V.Gubin}
 \affiliation{Institute of Laser Physics SB RAS, 630090, Novosibirsk, Russia}
 \author{K.V.Lotov}
 \affiliation{Budker Institute of Nuclear Physics SB RAS, 630090, Novosibirsk, Russia}
 \affiliation{Novosibirsk State University, 630090, Novosibirsk, Russia}
 \author{V.I.Trunov}
 \affiliation{Institute of Laser Physics SB RAS, 630090, Novosibirsk, Russia}
 \affiliation{Novosibirsk State University, 630090, Novosibirsk, Russia}
 \author{E.V.Pestryakov}
 \affiliation{Institute of Laser Physics SB RAS, 630090, Novosibirsk, Russia}
 \affiliation{Novosibirsk State University, 630090, Novosibirsk, Russia}
 \date{\today}
 \begin{abstract}
Powerful femtosecond laser pulses that propagate through narrow ablating capillaries cause modification of capillary walls, which is studied experimentally and theoretically. At low intensities, laser-induced periodic surface structures (LIPSS) and porous coating composed of sub-micron particles appear on the walls. At higher intensities, the surface is covered by deposited droplets of the size up to 10\,$\mu$m. In both cases, the ablated material forms a solid plug that completely blocks the capillary after several hundreds or thousands of pulses. The suggested theoretical model indicates that plug formation is a universal effect. It must take place in any narrow tube subject to ablation under the action of short laser pulses.
 \end{abstract}
 \maketitle

\section{Introduction}

The advent of high-power femtosecond lasers\cite{RMP78-309} gave impetus to development of novel plasma-based acceleration techniques that could offer new compact sources of high-energy particles or coherent radiation.\cite{RMP81-1229,PoP19-055501,RMP85-1,PPCF56-084015,RMP85-751,UFN57-1149} Certain of particle acceleration schemes use laser pulse guiding in plasma-filled capillaries. The capillaries serve for preventing diffraction of the pulse and extending the acceleration length either directly by reflecting the pulse from capillary walls,\cite{PRL82-4655,IEEE-PS28-1071,PRL92-205002,APB105-309,PoP20-083106,PRST-AB17-051302,PoP19-093121,PAc63-139,LPB19-219,PoP22-103111} or indirectly through a specific plasma profile inside.\cite{NatPhys2-696,PPCF49-B403,CRP10-130,PRL113-245002,APL99-091502,NJP9-415,NatPhys7-867,PoP16-123103,PoP16-093101} Capillary-guided laser pulses may find applications for X-ray lasing\cite{LPB9-725,PRA73-033801} and  high-harmonic generation.\cite{PRL83-2187} A novel concept of wakefield acceleration in hollow plasma channels,\cite{PoP7-3031,PoP20-080701,PoP20-123115,PRST-AB16-071301,SRep4-4171} which offers a number of advantages over usual plasma acceleration, may also stimulate interest in laser-capillary interaction. One of possible options for producing such a hollow channel could be capillary ablation under the action of a short laser pulse.

\begin{figure}[b]
\includegraphics[width=217bp]{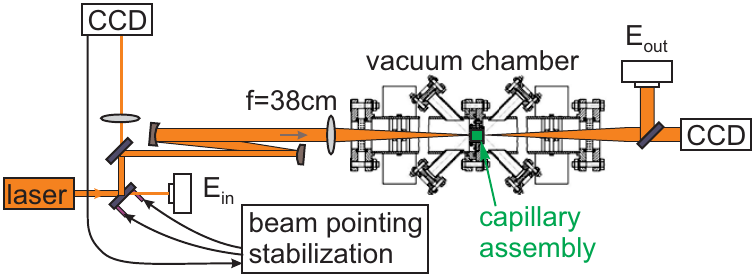}
\caption{ Experimental setup.}\label{fig1-setup}
\end{figure}

For the above-mentioned applications, it is important to understand how the capillary erodes under the action of many laser pulses. Because of a large ratio of the capillary length to the diameter, we might expect a slower erosion than that of open surfaces. The ablated material has a high probability to condense back to the walls when the capillary cools down between the pulses, and this effect gives promise of extending the capillary lifetime. The main process of interest is thus redistribution of the material along the capillary.

This paper is the extension of Ref.~\onlinecite{PoP22-103111}. We study modification of narrow capillaries under the influence of short powerful laser pulses both experimentally (Sec.~\ref{s2}) and theoretically (Sec.~\ref{s3}). We identify formation of solid plugs as the main effect responsible for capillary degradation. A universal character of this effect and a possible cure for this problem are discussed in Sec.~\ref{s4}.

\section{Experiments}\label{s2}

The experiments are performed at the Institute of Laser Physics SB RAS. The experimental setup is shown in Fig.\,\ref{fig1-setup}. The laser system\cite{LPL11-095301} delivers 50\,fs, up to 60\,mJ, pointing stabilized pulses at 850 nm central wavelength with a time contrast better than $10^{-8}$ to the capillary entrance. The capillary is rigidly fixed inside the vacuum chamber located on the five-dimensional adjustment stage. Input and output energies ($E_\text{in}$, $E_\text{out}$) are registered by pyroelectric energy meters. The spacial profile of the transmitted radiation is measured by a CCD camera located 68 cm from the capillary exit. The pulses follow at 10 Hz.
\begin{figure}[t]
\includegraphics[width=223bp]{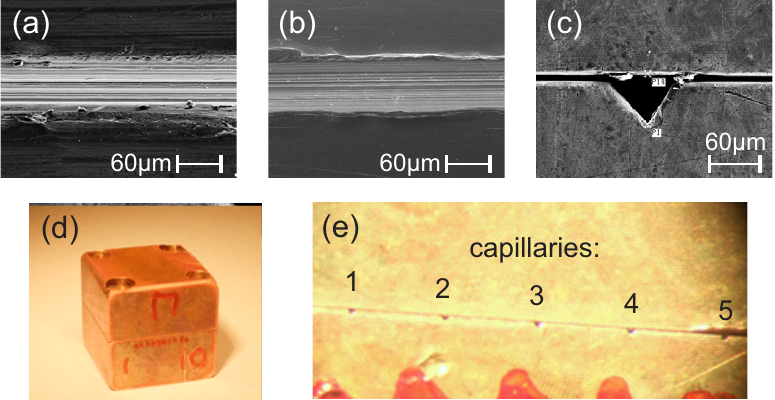}
\caption{ Scanning electron microscope images of capillary groves before (a) and after (b) polishing; (c) the entrance to the capillary (the black line between the plates is caused by small cants on the edges of the plates, the real gap is smaller than 0.5\,$\mu$m); (d) a capillary assembly, and (e) entry holes of capillaries.}\label{fig2-capillary}
\end{figure}
\begin{figure}[tb]
\includegraphics[width=227bp]{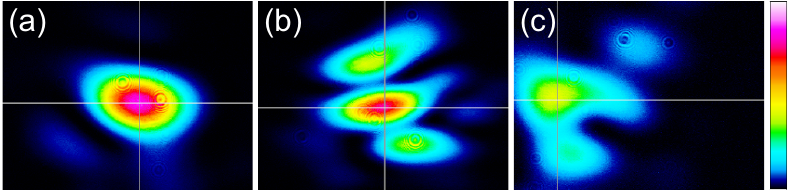}
\caption{ Far-field distributions of the transmitted radiation for (a) a good capillary and optimum entry conditions, (b) the same good capillary, but with a mismatched incident pulse, (c) a poor quality capillary. The pictures show spots in the rectangular angle frame 60\,mrad $\times$ 45\,mrad, the intensity (color map) is in proportional arbitrary units.}\label{fig3-spots}
\end{figure}
\begin{figure}[tb]
\includegraphics[width=232bp]{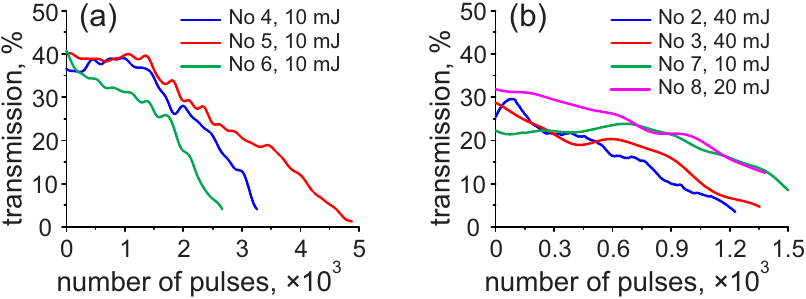}
\caption{ Typical dependencies of capillary transmission on the number of pulses for the pulse energy 10\,mJ (a) and for higher energies (b). ``No'' is a capillary number in an assembly.}\label{fig4-transmission}
\end{figure}
\begin{figure}[tb]
\includegraphics[width=230bp]{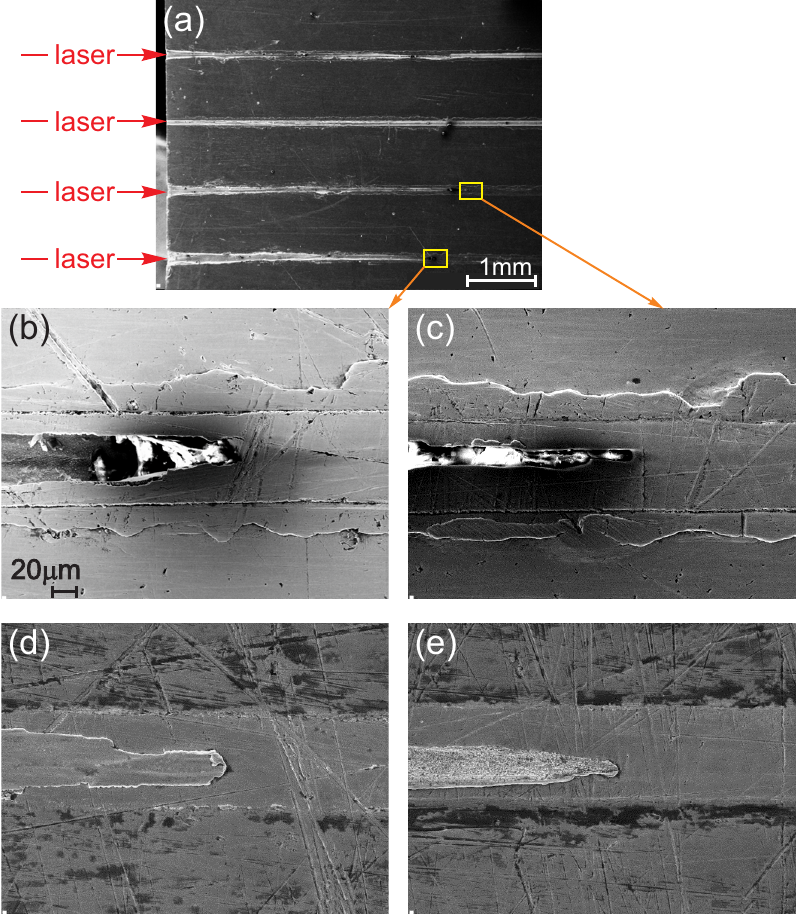}
\caption{ General view of plugged capillaries (a), and zoomed in plug areas: grooved (b),(c) and complementary (d),(e) plates. About 3500 pulses of the energy $\sim$10\,mJ were injected into these capillaries. }\label{fig5-plugs}
\end{figure}
\begin{figure}[tb]
\includegraphics[width=211bp]{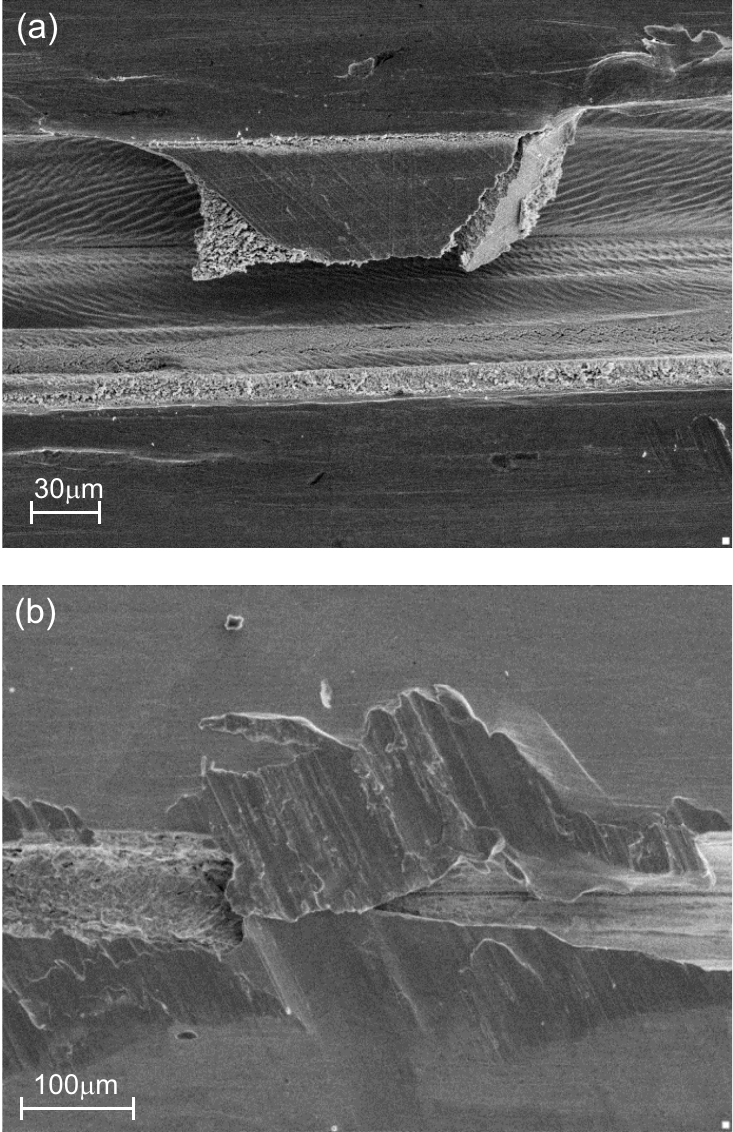}
\caption{ An incipient plugs on grooved plates for ``low'' (a) and ``high'' (b) pulse energies. The pulse energy (the number of pulses) is 2\,mJ (4000) in case (a) and 60\,mJ (40) in case (b).}\label{fig6-young}
\end{figure}
\begin{figure}[tb]
\includegraphics[width=230bp]{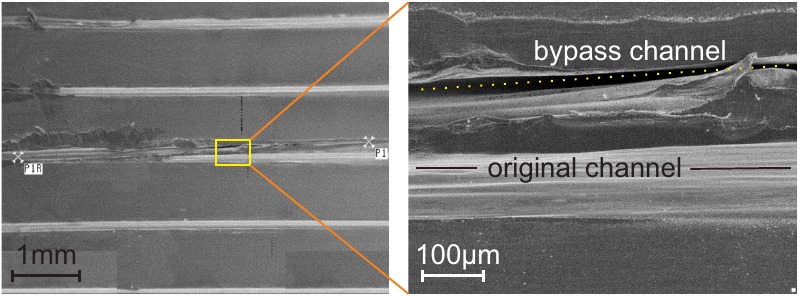}
\caption{ A bypass channel drilled by the beam. The pulse energy is about 10\,mJ; 6000 pulses were injected, about 500 pulses have passed. }\label{fig7-bypass}
\end{figure}

The capillaries are grooved on one side of a polished oxygen-free copper plate; another (not grooved) plate closes the capillary (Fig.\,\ref{fig2-capillary}). Each assembly (couple of plates with fastenings) contains up to 10 parallel capillaries. The capillaries selected as ``good'' have a nearly-equilateral triangular shape with the height (depth) $60\pm 5\,\mu$m, which is held to within $\pm 1.5\,\mu$m along the capillary length of 2\,cm. The capillary quality is controlled with a scanning electron microscope and by measuring capillary transmission at low intensities. The highest transmission (50--80\%) is observed for the focal spot (on the $e^{-2}$ level) of 0.6 times the channel depth. In this case, the transmitted radiation bears witness to single-mode pulse propagation [Fig.\,\ref{fig3-spots}(a)]. Poor capillary quality, mismatched spot size or wrong incidence angle result in multi-mode propagation and low transmission [Fig.\,\ref{fig3-spots}(b,c)]. In the reported experiments, the beam focusing was not always fine tuned for the highest transmission, but this is not essential for studied phenomena.

In all experiments, the capillary transmission decreases almost linearly with the number of passed pulses (Fig.\,\ref{fig4-transmission}). The reason for that was discovered on examining the used capillaries under a microscope: there are solid copper plugs formed at distances of 2--4\,mm from the entrance (Fig.\,\ref{fig5-plugs}). A plug initially appears as a local narrowing of the capillary (Fig.\,\ref{fig6-young}). If a grown plug is exposed to the laser beam for a long time, then the beam burns a dead-end bypass channel instead of destroying the plug (Fig.\,\ref{fig7-bypass}).

\begin{figure}[tb]
\includegraphics[width=232bp]{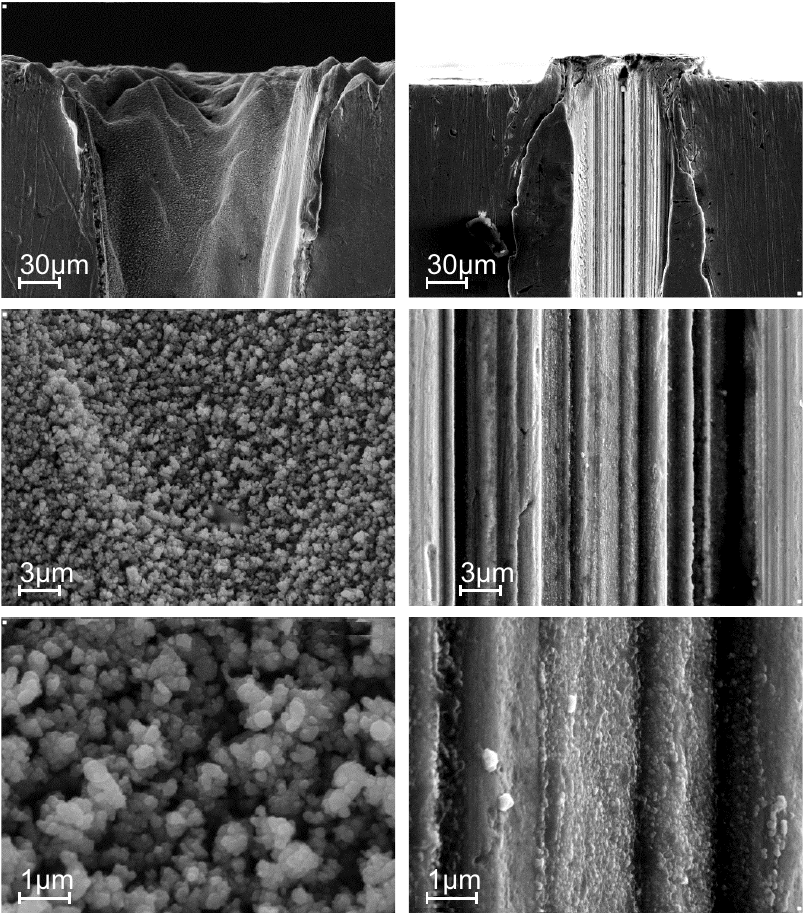}
\caption{ Capillary surface near the entrance (left) and exit (right) at different microscope resolutions for a ``low'' pulse energy. About 3000 pulses of the energy 10\,mJ have passed.}\label{fig8-porous}
\end{figure}

The surface structure of used capillaries bears similarities with that observed in ablation or drilling experiments. At low pulse energies, the so-called laser-induced periodic surface structures\cite{PRB26-5366,IEEE-QE22-1384,OptLett32-1932,PRB79-033409,ACSNano3-4062,CWLuo} (LIPSS) appear on capillary walls [Fig.\,\ref{fig6-young}(a)]. Unlike most of reported experiments, the rippling period in our experiments ($\sim$3\,$\mu$m) is several times larger than the laser wavelength, which may be due to waveguide-type propagation of the pulse. The modified capillary surface [Fig.\,\ref{fig8-porous}(left)] and the plugs [Fig.\,\ref{fig6-young}(a)] are porous and composed of submicron-size particles. This is rather unusual in view of the fact that porous structure of debris is typical for experiments in gas and not in vacuum.\cite{PRB72-195422,JPCS59-579,JAP108-034322,IJES7-1666} Experimental evidences of porous surface production in vacuum also exist in literature,\cite{APA90-537,APB119-445} though.

\begin{figure*}[tb]
\includegraphics[width=384bp]{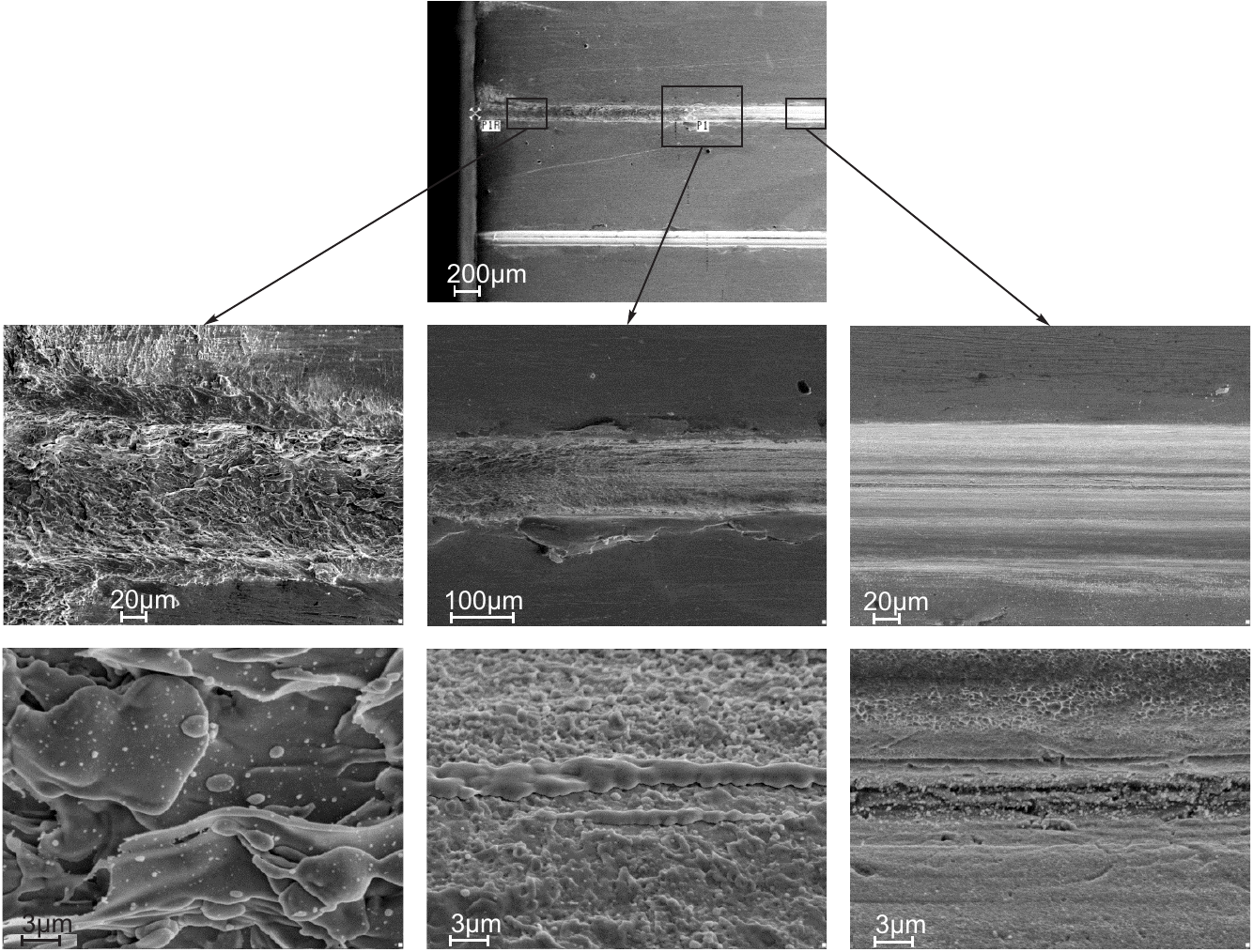}
\caption{ Capillary surface at various depths for a ``high'' pulse energy. About 20 pulses of the energy 45--60\,mJ have passed.}\label{fig9-droplets}
\end{figure*}
At higher pulse energies, the inner surface of capillaries is covered by deposited droplets (Fig.\,\ref{fig9-droplets}). The droplet size depends on the local energy density in the particular place and varies between $\sim$100\,nm and $\sim$10\,$\mu$m. No large droplets ($>1\,\mu$m) appear if the pulse energy is below 20\,mJ (Fig.\,\ref{fig8-porous}). The transition between areas of large and small droplets is rather sharp and located at the typical depth of plug formation (Fig.\,\ref{fig9-droplets}). And vice versa, the plugs separate regions of large and small droplets [Fig.\,\ref{fig6-young}(b)].

\begin{figure}[tb]
\includegraphics[width=232bp]{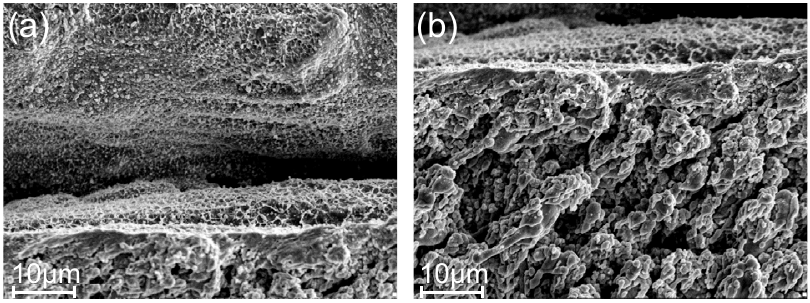}
\caption{ The ``cotton-like'' structure of capillary walls at the place of narrowing (a) and of a plug (b).  About 1500 pulses of the energy 11\,mJ have passed in helium at 100\,mbar.}\label{fig10-cotton}
\end{figure}
\begin{figure}[tb]
\includegraphics[width=233bp]{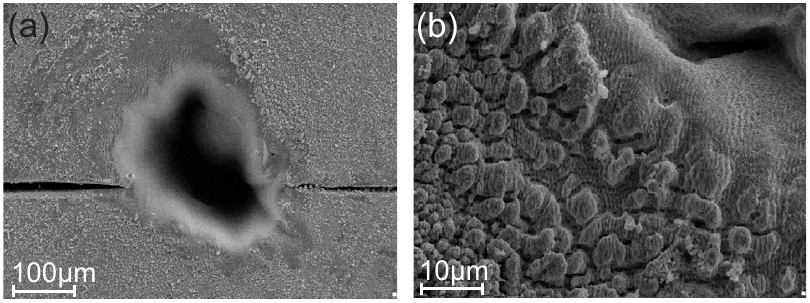}
\caption{ The coating around a capillary entrance. About 2500 pulses of the energy $10 \div 15$\,mJ (12.5\,mJ on average) have passed in helium at 100\,mbar. Both snapshots are made near the same capillary, but with different magnifications.}\label{fig11-black}
\end{figure}

Similar experiments carried out by us in helium at the pressure interval $5\times 10^{-3} \div 1200$ mbar show no measurable differences in capillary transmission and capillary lifetime. The only differences are in surface modifications. A ``cotton-like'' structure\cite{PRB72-195422,JPCS59-579,APA90-537} (bridges between the droplets) appears on capillary walls and inside the plugs (Fig.\,\ref{fig10-cotton}). The entrance holes are surrounded by copper droplets that act as an absorbing coating and look like black halo around the holes (Fig.\,\ref{fig11-black}). This is in line with experiments that show the effect of the surrounding gas on droplet formation\cite{IJES7-1666,JAP108-034322} and low reflectivity of the formed coatings.\cite{PRB72-195422,JPCS59-579,JAP117-033103}

\begin{figure*}[tb]
\includegraphics[width=470bp]{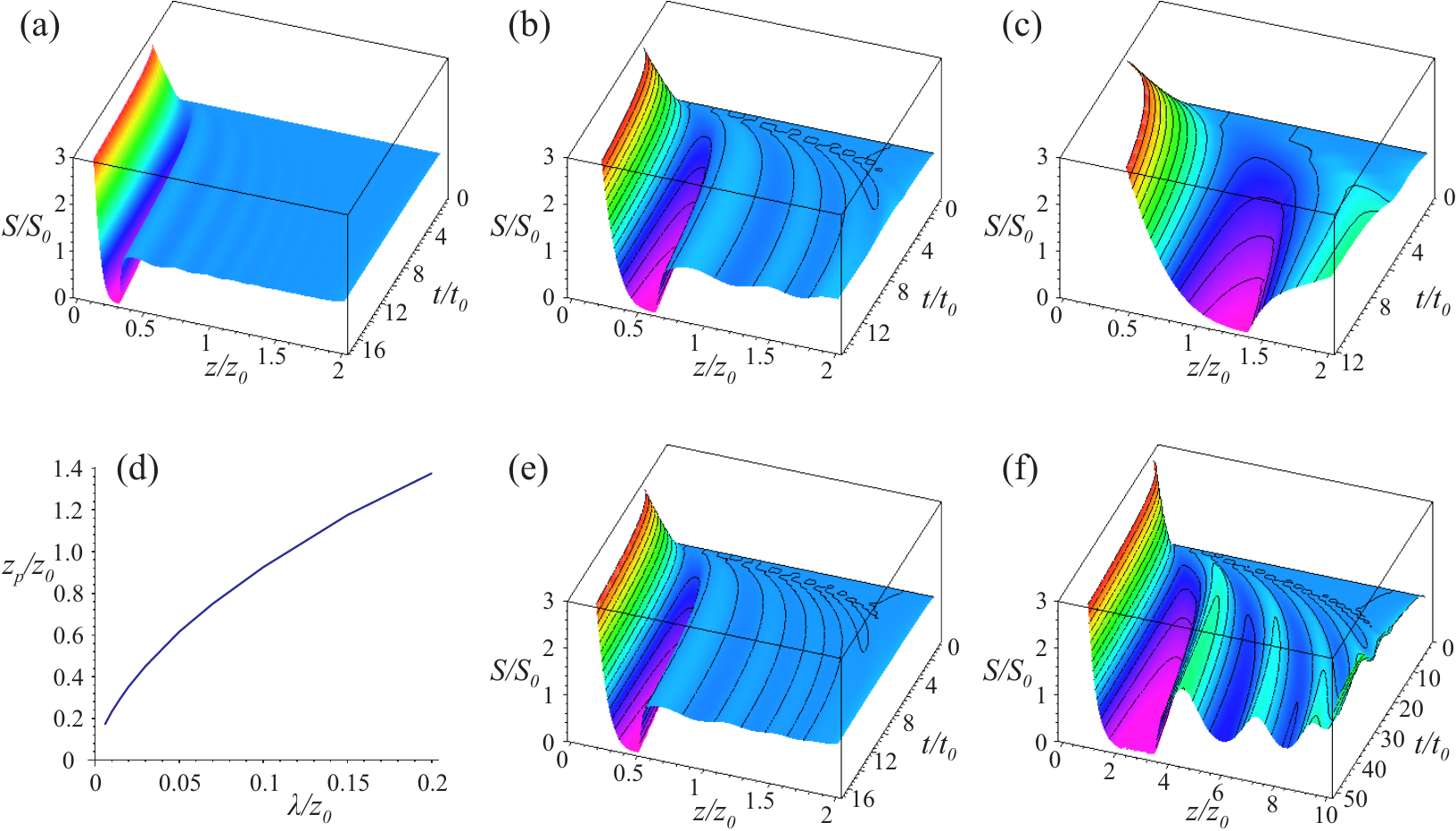}
\caption{ Calculated capillary cross-sections for the material distribution law (\ref{e6}), capillary length $L=2 z_0$, and $\lambda=0.015 z_0$ (a), $\lambda=0.05 z_0$ (b), and $\lambda=0.2 z_0$ (c); dependence of the plug location $z_p$ on the parameter $\lambda$ (d); the calculated capillary cross-section for the material distribution law (\ref{e7}), $L=2 z_0$, and $\lambda=0.05 z_0$ (e), and for the material distribution law (\ref{e7}), $L=10 z_0$, and $\lambda=0.2 z_0$ (f). }\label{fig12-theory}
\end{figure*}

\section{Theory}\label{s3}

In this Section, we develop a theoretical model that will give us an insight into which effects may be responsible for plug formation. We assume that the role of the laser pulse consists in instant heating of capillary walls and ablation of a certain amount of material. We also take into account that energy deposition along the capillary depends on the capillary cross-section, and the ablated material redistributes along the capillary. With these effects in the basis, we find how does the capillary cross-section change as a large number of laser pulses passes through.

Denote the distance along the capillary by $z$, the time by $t$, the capillary length by $L$, the energy of the laser pulse by $W(z,t)$, the capillary cross-section by $S(z,t)$, and the energy deposited by the pulse in a unit length of a capillary by $Q(z,t)$. All pulses have the same energy $W_0$ at the point of entry at $z=0$.

If the cross-section is constant along the capillary, then the pulse energy decreases exponentially with some damping distance $z_0$:
\begin{equation}\label{e2}
    \frac{\partial W}{\partial z} = - \frac{W}{z_0}.
\end{equation}
If the capillary cross-section sharply decreases, the pulse energy decreases also. We assume $W/S = \const$ in this place, which in the differential form reads as
\begin{equation}\label{e3}
    \frac{\partial W}{\partial z} = \frac{W}{S} \frac{\partial S}{\partial z}.
\end{equation}
Let us combine effects (\ref{e2}) and (\ref{e3}) into a single equation. It will be an approximate one, but still suitable for qualitative studies. The pulse energy cannot grow along the capillary, so we introduce a logical operator into the equation:
\begin{equation}\label{e4}
    \frac{\partial W}{\partial z} = -Q = -\max \left( \frac{W}{z_0} - \frac{W}{S} \frac{\partial S}{\partial z}, \quad 0 \right).
\end{equation}

Then we assume that the mass of the ablated material is proportional to the local energy release $Q$. After a short period of time, the capillary cools down, and the material returns back to the walls. As a result, the cross-section changes:
\begin{equation}\label{ne5}
    \frac{\partial S}{\partial t} = A (Q - Q_a),
\end{equation}
where the constant $A$ accounts for the ablation rate, periodicity of pulses, and geometrical factors. The quantity $Q_a$ is proportional to the amount of re-deposited material and, for convenience, is measured in the same units as $Q$. Assume also that the law of material distribution is the same in all cross-sections and is determined by the convolution integral:
\begin{equation}\label{ne6}
    Q_a (z) = \int_0^L Q(z') f(z-z') \, dz', \quad \int_{-\infty}^\infty f(x) \, dx = 1.
\end{equation}
The exact shape of the function $f(x)$ and the scale of its variation may affect the result, so we consider two variants:
\begin{equation}\label{e6}
    f(x) = \frac{1}{\sqrt{2 \pi \lambda}} \exp \left( -\frac{x^2}{2 \lambda^2} \right),
\end{equation}
\begin{equation}\label{e7}
    f(x) = \begin{cases}
    (2\lambda)^{-1}, \qquad & |x|<\lambda, \\
    0, & \text{otherwise},
    \end{cases}
\end{equation}
and vary $\lambda$ to determine sensitivity of our model to this parameter.

If we take $W_0$, $z_0$, and the initial cross-section $S_0$ as units of measure for corresponding quantities, $W_0/z_0$ as the unit for $Q$ and $Q_a$, and
\begin{equation}\label{ne9}
    t_0 = \frac{S_0 z_0}{AI_0}
\end{equation}
as the time unit, then equations (\ref{e4}) and (\ref{ne5}) take the universal form:
\begin{gather}
\label{ne10}
    \frac{\partial W}{\partial z} = -Q = -\max \left( W - \frac{W}{S} \frac{\partial S}{\partial z}, \quad 0 \right), \\
\label{ne11}
    \frac{\partial S}{\partial t} = Q - Q_a
\end{gather}
with initial (boundary) conditions
\begin{equation}\label{ne12}
    W(0,t) = 1, \qquad S(z,0)=1.
\end{equation}

The solution of equations (\ref{ne10})--(\ref{ne12}) and (\ref{ne6})--(\ref{e7}) depends only on the ratio of material expansion distance $\lambda$, pulse damping length $z_0$ and capillary length $L$. Typical solutions are shown in Fig.\,\ref{fig12-theory}. In the cases of interest ($L \gtrsim z_0 \gg \lambda$), the capillary cross-section always goes down to zero at some place, that is, a plug is formed. The location of the plug $z_p$ depends on the material expansion distance $\lambda$, the smaller the closer to the entrance [Fig.\,\ref{fig12-theory}(d)]. The result is insensitive to the choice of the material expansion law  $f(x)$, as is seen from comparison of Figs.\,\ref{fig12-theory}(b) and~\ref{fig12-theory}(e). If the capillary is long compared to both $\lambda$ and $z_0$, and the ratio $\lambda/z_0$ is moderately small, then multiple narrow spots develop in the capillary, first of which becomes the plug [Fig.\,\ref{fig12-theory}(f)]. At lower ratios $\lambda/z_0$, a small-amplitude rippling is also visible [Fig.\,\ref{fig12-theory}(b),(e)]. Therefore the plug growth can be considered as development of some instability and characterized by period and growth rate of the most unstable perturbation.

In the considered theoretical model, all parameters that are difficult to evaluate affect only the timescale $t_0$ of the process through the constant $A$. The distances can be easily measured and compared to the model; this gives an estimate for $\lambda$. In the experiments we have $L=2$\,cm, $z_p \sim 0.3$\,cm, and $z_0 \approx 1$\,cm; the latter corresponds to the measured 60\% transmission through the capillary. Then we obtain from Fig.\,\ref{fig12-theory}(d) that $\lambda \sim 0.015z_0 \sim 150\,\mu$m. This is close to the capillary width and seems a reasonable value in view of the observation that the material is transferred in the form of small droplets and therefore propagate along straight lines after ablation. The measured plug length ($\sim$1\,mm) is roughly 1/3 of $z_p$, which also agrees with the theory. The experiments thus correspond to the case shown in Fig.\,\ref{fig12-theory}(a).

\section{Discussion}\label{s4}

There is a simple physical effect that underlies plug formation. The material travels in both directions along the capillary. Roughly half of the ablated material moves forward and sees a week oncoming stream, as there is almost no ablation in downstream areas. This unbalanced material flow (permanently supported by the repeating pulses) results in material accumulation at some place and formation of a plug there. Note that the model predicts plug formation even with no account of specific volume increase. In the experiments we see that the re-deposited material is porous and thus has a larger volume, which additionally favors capillary blocking.

Since the plugs are a consequence of basic effects, they may appear in various setups wherever multiple laser pulses cause material ablation inside a narrow tube. One such example is laser drilling of high aspect ratio channels. It was discovered that there exists a limit for the channel depth, and the channel is often branched in its deepest part.\cite{OptExp20-27147,APA112-623,SPIE-8247-824717} The reported aspect ratio for drilling experiments with a similar pulse radius\cite{OptExp20-27147,SPIE-8247-824717} is about 25:1, which is close to the ratio of the plug coordinate ($z_p \sim 2-4$\,mm) to the final channel diameter at the entrance ($\sim$150$\,\mu$m, Fig.\,\ref{fig8-porous}) in our experiments. Both the depth limit and the branching may be explained by the plugs. Once a plug appears at a certain depth, the laser drills a bypass channel similar to that shown in Fig.\,\ref{fig7-bypass}. The bypass channel eventually gets blocked with another plug, and so on.

The mechanism of plug formation suggests a key to get rid of the plug problem. The uncompensated inward flow of the material is caused by the increased ablation rate in areas where the capillary cross-section decreases. This seems unavoidable if the ablation is caused by laser pulses. However, if the ablation would increase in areas of small cross-section, the incipient plugs would erode away. In other words, the ablation rate must inversely depend on the capillary cross-section $S$ rather than scale as the derivative $\partial S/\partial z$. This feature may be realized in "healing" capillary discharges that would flatten the capillary after a number of laser pulses, but this option needs further studies.

To conclude, we experimentally studied modifications of narrow ablating capillaries under the influence of multiple femtosecond laser pulses. We discovered the effect of solid plug formation inside the capillaries, described the observed plugs, and suggested a theoretical model of plug growth. Predictions of the model agrees with experiments.

\acknowledgements

This work is partly supported by the RAS Program of Basic Research ``Extreme Laser Radiation: Physics and Fundamental Applications''. The authors are grateful to Ya.\,L.\,Lukyanov for microscope image taking and to V.V.Ershov for capillary design and technology development.


\begin{thebibliography}{88}
  \bibitem{RMP78-309}
	G.A.Mourou, T.Tajima, and S.V.Bulanov
   	Rev. Mod. Phys. \textbf{78}, 309 (2006).
  \bibitem{RMP81-1229}
   E. Esarey, C. B. Schroeder, and W. P. Leemans,
   Rev. Mod. Phys. \textbf{81}, 1229 (2009).
\bibitem{PoP19-055501}
    V.Malka,
    Phys. Plasmas {\bf 19}, 055501 (2012).
 \bibitem{RMP85-1}
    S. Corde, K. Ta Phuoc, G. Lambert, R. Fitour, V. Malka, A. Rousse, A. Beck, and E. Lefebvre,
    Rev. Mod. Phys. {\bf 85}, 1 (2013).
\bibitem{PPCF56-084015}
    F.Albert, A.G.R.Thomas, S.P.D.Mangles, S.Banerjee, S.Corde, A.Flacco, M.Litos, D.Neely, J.Vieira, Z.Najmudin, R.Bingham, C.Joshi, and T.Katsouleas,
    Plasma Phys. Control. Fusion \textbf{56}, 084015 (2014).
\bibitem{RMP85-751}
    A.Macchi, M.Borghesi, M.Passoni,
    Rev. Mod. Phys. \textbf{85}, 751 (2013).
\bibitem{UFN57-1149}
    S.V. Bulanov, J.J. Wilkens, T.Zh. Esirkepov, G. Korn, G. Kraft, S.D. Kraft, M. Molls, V.S. Khoroshkov,
    Physics -- Uspekhi \textbf{57}, 1149 (2014).
 \bibitem{PRL82-4655}
    F.Dorchies, J.R.Marques, B.Cros, G.Matthieussent, C.Courtois, T.Velikoroussov, P.Audebert, J.P.Geindre, S.Rebibo, G.Hamoniaux, and F.Amiranoff,
    Phys. Rev. Lett. \textbf{82}, 4655 (1999).
 \bibitem{IEEE-PS28-1071}
    B.Cros, C.Courtois, G.Malka, G.Matthieussent, J.R.Marques, F.Dorchies, F.Amiranoff, S.Rebibo, G.Hamoniaux, N.Blanchot, and J.L.Miquel,
    IEEE Trans. Plasma Sci. \textbf{28}, 1071 (2000).
 \bibitem{PRL92-205002}
    Y.Kitagawa, Y.Sentoku, S.Akamatsu, W.Sakamoto, R.Kodama, K.A.Tanaka, K.Azumi, T.Norimatsu, T.Matsuoka, H.Fujita, and H.Yoshida,
    Phys. Rev. Lett. \textbf{92}, 205002 (2004).
\bibitem{APB105-309}
    G.Genoud, K.Cassou, F.Wojda, H.E.Ferrari, C.Kamperidis, M.Burza, A.Persson, J.Uhlig, S.Kneip, S.P.D.Mangles, A.Lifschitz, B.Cros, C.-G.Wahlstr\"om,
    Appl. Phys. B \textbf{105}, 309 (2011).
 \bibitem{PoP20-083106}
    J.Ju, K.Svensson, H.Ferrari, A.D\"opp, G.Genoud, F.Wojda, M.Burza, A.Persson, O.Lundh, C.-G.Wahlstr\"om, and B.Cros,
    Phys. Plasmas \textbf{20}, 083106 (2013).
 \bibitem{PRST-AB17-051302}
    J.Ju, G.Genoud, H.E.Ferrari, O.Dadoun, B.Paradkar, K.Svensson, F.Wojda, M.Burza, A.Persson, O.Lundh, N.E.Andreev, C.-G.Wahlstr\"om, and B.Cros,
    Phys. Rev. ST Accel. Beams \textbf{17}, 051302 (2014).
 \bibitem{PoP19-093121}
    V.Eremin, Yu.Malkov, V.Korolikhin, A.Kiselev, S.Skobelev, A.Stepanov, and N.Andreev,
    Phys. Plasmas \textbf{19}, 093121 (2012).
 \bibitem{PAc63-139}
    K.V.Lotov,
    Part. Accel. \textbf{63}, 139 (1999).
 \bibitem{LPB19-219}
    K.V.Lotov,
    Laser Part. Beams \textbf{19}, 219 (2001).
\bibitem{PoP22-103111}
	K.V.Lotov, K.V.Gubin, V.E.Leshchenko, V.I.Trunov, and E.V.Pestryakov,
	Phys. Plasmas \textbf{22}, 103111 (2015).
  \bibitem{NatPhys2-696}
    W.P.Leemans, B.Nagler, A.J.Gonsalves, Cs.Toth, K.Nakamura, C.G.R.Geddes, E.Esarey, C.B.Schroeder, and S.M.Hooker,
    Nat. Phys. \textbf{2}, 696 (2006).
\bibitem{PPCF49-B403}
    S.M. Hooker, E. Brunetti, E. Esarey, J.G. Gallacher, C.G.R. Geddes, A.J. Gonsalves, D.A. Jaroszynski, C. Kamperidis, S. Kneip, K. Krushelnick, W.P. Leemans, S.P.D. Mangles, C.D Murphy, B. Nagler, Z. Najmudin, K. Nakamura, P.A. Norreys, D. Panasenko, T.P. Rowlands-Rees, C.B. Schroeder, Cs. Toth, and R. Trines,
    Plasma Phys. Control. Fusion \textbf{49}, B403 (2007).
\bibitem{CRP10-130}
    W.P. Leemans, E. Esarey, C.G.R. Geddes, Cs. Toth, C.B. Schroeder, K. Nakamura, A.J. Gonsalves, D. Panasenko, E. Cormier-Michel, G.R. Plateau, C. Lin, D.L. Bruhwiler, J.R. Cary,
    C. R. Physique \textbf{10}, 130 (2009).
 \bibitem{PRL113-245002}
    W.P.Leemans, A.J.Gonsalves, H.-S.Mao, K.Nakamura, C.Benedetti, C.B.Schroeder, Cs.Toth, J.Daniels, D.E.Mittelberger, S.S.Bulanov, J.-L.Vay, C.G.R.Geddes, and E.Esarey,
    Phys. Rev. Lett. \textbf{113}, 245002 (2014).
\bibitem{APL99-091502}
    H. Lu, M. Liu, W. Wang, C. Wang, J. Liu, A. Deng, J. Xu, C. Xia, W. Li, H. Zhang, X. Lu, C. Wang, J. Wang, X. Liang, Y. Leng, B. Shen, K. Nakajima, R. Li, and Z. Xu,
    Appl. Phys. Lett. \textbf{99}, 091502 (2011).
\bibitem{NJP9-415}
    S. Karsch, J. Osterhoff, A. Popp, T.P. Rowlands-Rees, Zs. Major, M. Fuchs, B. Marx, R. Horlein, K. Schmid, L. Veisz, S. Becker, U. Schramm, B. Hidding, G. Pretzler, D. Habs, F. Gruner, F. Krausz and S.M. Hooker,
    New Journal of Physics \textbf{9}, 415 (2007).
\bibitem{NatPhys7-867}
    S. Cipiccia, M.R. Islam, B. Ersfeld, R.P. Shanks, E. Brunetti, G. Vieux, X. Yang, R.C. Issac, S.M. Wiggins, G.H. Welsh, M.-P. Anania, D. Maneuski, R. Montgomery, G. Smith, M. Hoek, D.J. Hamilton, N.R.C. Lemos, D. Symes, P.P. Rajeev, V.O. Shea, J.M. Dias, and D.A. Jaroszynski,
    Nat. Phys. \textbf{7}, 867 (2011).
\bibitem{PoP16-123103}
    Y. Mori, Y. Sentoku, K. Kondo, K. Tsuji, N. Nakanii, S. Fukumochi, M. Kashihara, K. Kimura, K. Takeda, K. A. Tanaka, T. Norimatsu, T. Tanimoto, H. Nakamura, M. Tampo, R. Kodama, E. Miura, K. Mima, and Y. Kitagawa,
    Phys. Plasmas \textbf{16}, 123103 (2009).
\bibitem{PoP16-093101}
    T. Kameshima, H. Kotaki, M. Kando, I. Daito, K. Kawase, Y. Fukuda, L. M. Chen, T. Homma, S. Kondo, T. Zh. Esirkepov, N.A. Bobrova, P.V. Sasorov, and S.V. Bulanov,
    Phys. Plasmas \textbf{16}, 093101 (2009).
\bibitem{LPB9-725}
    C.St\"ockl, G.D.Tsakiris,
    Laser Part. Beams \textbf{9}, 725 (1991).
 \bibitem{PRA73-033801}
    B.Cros, T.Mocek, I.Bettaibi, G.Vieux, M.Farinet, J.Dubau, S.Sebban, and G.Maynard,
    Phys. Rev. A \textbf{73}, 033801 (2006).
 \bibitem{PRL83-2187}
    C.G.Durfee III, A.R.Rundquist, S.Backus, C.Herne, M.M.Murnane, and H.C.Kapteyn,
    Phys. Rev. Lett. \textbf{83}, 2187 (1999).

  \bibitem{PoP7-3031}
	B.Rau and R.A.Cairns,
	Phys. Plasmas \textbf{7}, 3031 (2000).
\bibitem{PoP20-080701}
    C.B. Schroeder, E. Esarey, C. Benedetti, and W.P. Leemans,
    Phys. Plasmas \textbf{20}, 080701 (2013).
\bibitem{PoP20-123115}
    C.B. Schroeder, C. Benedetti, E. Esarey, and W.P. Leemans,
    Phys. Plasmas \textbf{20}, 123115 (2013).
 \bibitem{PRST-AB16-071301}
    L.Yi, B.Shen, K.Lotov, L.Ji, X.Zhang, W.Wang, X.Zhao, Y.Yu, J.Xu, X.Wang, Y.Shi, L.Zhang, T.Xu, Z.Xu,
    Phys. Rev. ST Accel. Beams \textbf{16}, 071301 (2013).
 \bibitem{SRep4-4171}
	L.Yi, B.Shen, L.Ji, K.Lotov, A.Sosedkin, X.Zhang, W.Wang, J.Xu, Y.Shi, L.Zhang, and Z.Xu,
	Scientific Reports \textbf{4}, 4171 (2014).
 \bibitem{LPL11-095301}
    V.E.Leshchenko, V.I.Trunov, S.A.Frolov, E.V.Pestryakov, V.A.Vasiliev, N.L.Kvashnin, and S.N.Bagayev,
    Laser Phys. Lett. \textbf{11}, 095301 (2014).

\bibitem{PRB26-5366}
    Z. Guosheng, P.M. Fauchet, and A.E. Siegman,
    Phys. Rev. B \textbf{26}, 5366 (1982).
\bibitem{IEEE-QE22-1384}
    A.E. Siegman, P.M. Fauchet,
    IEEE J. Quant. Electronics \textbf{QE-22}, 1384 (1986).
\bibitem{OptLett32-1932}
    Q.Z. Zhao, S. Malzer, and L.J. Wang,
    Opt. Lett. \textbf{32}, 1932 (2007).
\bibitem{PRB79-033409}
    S. Sakabe, M. Hashida, S. Tokita, S. Namba, and K. Okamuro,
    Phys. Rev. B \textbf{79}, 033409 (2009).
\bibitem{ACSNano3-4062}
    M. Huang, F. Zhao, Y. Cheng, N. Xu, and Z. Xu,
    ACS Nano \textbf{3}, 4062 (2009).
\bibitem{CWLuo}
    C.W.Luo,
    In: Lasers - Applications in Science and Industry, Dr Krzysztof Jakubczak (Ed.), ISBN: 978-953-307-755-0, InTech, 2011, p.3--22.

\bibitem{PRB72-195422}
    A. Y. Vorobyev and C. Guo,
    Phys. Rev. B \textbf{72}, 195422 (2005).
\bibitem{JPCS59-579}
    A.Y. Vorobyev and C.Guo,
    Journal of Physics: Conference Series \textbf{59}, 579 (2007).
\bibitem{JAP108-034322}
    D. Riabinina, E. Irissou, B. Le Drogoff, M. Chaker, and D. Guay,
    J. Appl. Phys. \textbf{108}, 034322 (2010).
\bibitem{IJES7-1666}
    Z. Hamoudi, M.A. El Khakani and M. Mohamedi,
    Int. J. Electrochem. Sci. \textbf{7}, 1666 (2012).
\bibitem{APA90-537}
    A. Weck, T.H.R. Crawford, D.S. Wilkinson, H.K. Haugen, J.S. Preston,
    Appl. Phys. A \textbf{90}, 537 (2008).
\bibitem{APB119-445}
    D.K. Pallotti, X. Ni, R. Fittipaldi, X. Wang, S. Lettieri, A. Vecchione, S. Amoruso,
    Appl. Phys. B \textbf{119}, 445 (2015).
\bibitem{JAP117-033103}
    A.Y. Vorobyev and C. Guo,
    J. Appl. Phys. \textbf{117}, 033103 (2015).

\bibitem{OptExp20-27147}
    S. D\"oring, J. Szilagyi, S. Richter, F. Zimmermann, M. Richardson, A. T\"unnermann, S. Nolte,
    Optics Express \textbf{20}, 27147 (2012).
\bibitem{APA112-623}
    S. D\"oring, S. Richter, F. Heisler, T. Ullsperger, A. T\"unnermann, S. Nolte,
    Appl. Phys. A \textbf{112}, 623 (2013).
\bibitem{SPIE-8247-824717}
    S. D\"oring, S. Richter, A. T\"unnermann, S. Nolte,
    Proc. of SPIE \textbf{8247}, 824717 (2012).

\end{thebibliography}
\end{document}